\documentclass[useAMS,usenatbib]{mn2e}
\usepackage{amsmath}
\usepackage{graphicx}
\title[Pulsational instability of supergiant protostars]
{Pulsational instability of supergiant protostars:\\
Do they grow supermassive by accretion? }
\author[K. Inayoshi, T. Hosokawa, and K. Omukai]{Kohei Inayoshi$^{1}$
\thanks{E-mail: inayoshi@tap.scphys.kyoto-u.ac.jp}, 
Takashi Hosokawa$^{2,3}$\thanks{E-mail: takashi.hosokawa@phys.s.u-tokyo.ac.jp}, 
and Kazuyuki Omukai$^{1}$
\thanks{E-mail: omukai@tap.scphys.kyoto-u.ac.jp}
\\
$^{1}$Department of Physics, Graduate School of Science, Kyoto University, 
Kyoto 606-8502, Japan \\
$^{2}$Department of Physics, University of Tokyo, Tokyo 113-0033, Japan \\
$^{3}$Jet Propulsion Laboratory, California Institute of Technology, 
Pasadena CA 91109, USA}

\newcommand{\msunyr}{{\rm M}_{\sun}~{\rm yr}^{-1}}
\newcommand{\mdot}{\dot{M}_{\rm acc}}
\newcommand{\msun}{{\rm M}_{\sun}}

\begin{document}

\date{\today}

\pagerange{000--000} \pubyear{0000}

\maketitle

\label{firstpage}

\begin{abstract}
Supermassive stars (SMSs; $M_\ast \ga 10^5~\msun$) and their remnant 
black holes are promising progenitors for supermassive black holes 
(SMBHs) observed in the early universe at $z\ga 7$. 
It has been postulated that SMSs forms through very rapid mass
accretion onto a protostar at a high rate exceeding $0.01~\msunyr$.
According to recent studies, such rapidly accreting protostars evolve
into ``supergiant protostars'', i.e. protostars consisting of 
a bloated envelope and a contracting core, similar to giant star.
However, like massive stars as well as giant stars, both of 
which are known to be pulsationally unstable, 
supergiant protostars may also be also unstable 
to launch strong pulsation-driven outflows. 
If this is the case, the stellar growth via accretion 
will be hindered by the mass loss.
We here study the pulsational stability of the supergiant protostars
in the mass range $M_\ast \la 10^3~\msun$
through the method of the linear perturbation analysis.
We find that the supergiant protostars with $M_\ast \ga 600~\msun$ 
and very high accretion rate $\mdot \ga 1.0~\msunyr$ are unstable 
due to the $\kappa$ mechanism.
The pulsation is excited in the He$^+$ ionization layer in the 
envelope.
Even under a conservative assumption that all the pulsation energy 
is converted into the kinetic energy of the outflows,
the mass-loss rate is $\sim 10^{-3}~\msunyr$, which is lower than 
the accretion rate by more than two orders of magnitude.
We thus conclude that the supergiant protostars should grow stably 
via rapid accretion at least in the mass range we studied.
As long as the rapid accretion is maintained in the later stage, 
protostars will become SMSs, 
which eventually produce seeds for the high-$z$ SMBHs.
\end{abstract}


%%%%%%%%%%%%%%%%%%%%%%%
\section{Introduction}
%%%%%%%%%%%%%%%%%%%%%%%

Recent observations of high-$z$ quasars reveal that
supermassive black holes (SMBHs) of 
$M_{\rm BH} \ga 10^9~\msun$ have
already formed as early as the beginning 
of the universe $\la$ 0.8Gyr 
(e.g., Fan 2006; Willott et al. 2007).
A popular formation scenario of those SMBHs postulates 
that remnant stellar-mass BHs 
($M_{\rm seed}\sim 100~\msun$) of Population III (Pop III) stars 
grow in mass via continuous mass accretion and merge
(e.g., Haiman \& Loeb 2001; Volonteri, Haardt \& Madau 2003; Li et al. 2007).
Given that the seed BHs grow at the Eddington mass-accretion rate
$\dot M_{\rm Edd}=L_{\rm Edd}/\epsilon c^2$, where $L_{\rm Edd}$ 
is the Eddington luminosity, and $\epsilon \simeq 0.1$ is 
the radiative efficiency,
the growth time to $10^9~\msun$ BHs is 
$\sim 0.05\ln(M_{\rm BH}/M_{\rm seed})$ Gyr $\simeq 0.8$ Gyr.
Since this growth time is as long as the age of the universe at
$z \simeq 7$, where the most distant SMBH is observed 
(Mortlock et al. 2011), the stellar-mass seed is required to keep
growing at least with the Eddington rate.
However, recent studies show that this is unlikely as the
accretion onto the BH, as well as onto the surrounding 
disk, is easily quenched by strong radiative feedback from 
the growing BH itself 
(Johnson \& Bromm 2007; Milosavljevi{\'c}, Couch \& Bromm 2009; 
Alvarez, Wise \& Abel 2009; Park \& Ricotti 2011; Park \& Ricotti 2012; 
Tanaka, Perna \& Haiman 2012).

%---------------------------------------------------------------------------%

In an alternative scenario, formation of supermassive stars 
(SMSs; $M_\ast \ga 10^5~\msun$) and their subsequent collapse 
directly to the BHs in the first galaxies 
($z \ga 10$, $T_{\rm vir} \ga 10^4$~K) has been envisaged
(e.g., Bromm \& Loeb 2003; Begelman, Volonteri \& Rees 2006;
Lodato \& Natarajan 2006). 
Here, primordial-gas clouds more massive than 
$10^5~\msun$ are supposed to contract monolithically 
to form stars without strong fragmentation.
Since rapid H$_2$ cooling causes fragmentation of the
primordial gas cloud, 
for the SMS formation, suppression of H$_2$ formation is required 
by some means. Examples of such means are: the photodissociation 
by far ultraviolet (FUV) radiation from nearby stars (e.g., Omukai 2001; Bromm \& Loeb
2003; Omukai, Schneider \& Haiman 2008; Regan \& Haehnelt 2009a,b;
Shang, Bryan, \& Haiman 2010; Inayoshi \& Omukai 2011; Agarwal et a. 2012; 
Johnson, Dalla Vecchia \& Khochfar 2012),
and the collisional dissociation in dense and hot gas 
(Inayoshi \& Omukai 2012).
The latter situation can be realized, for example, by the 
cold-accretion-flow shocks in the first galaxy formation.
In both cases, the primordial gas 
collapses isothermally at $T \simeq 8000$ K via H atomic cooling
(Ly$\alpha $, two-photon, and H$^-$ free-bound emission; Omukai 2001) and
no major fragmentation is observed in numerical simulations 
during this phase (Bromm \& Loeb 2003; Regan \& Haehnelt 2009a, b).

%----------------------------------------------------------------------------%

This monolithic contraction of the cloud leads to 
a formation of a small protostar ($\sim 0.01~\msun$) 
at its center. The embryo protostar subsequently 
grows to a SMS via rapid accretion of the surrounding envelope.
This process sounds similar to the case of ordinary 
Pop III star formation, where H$_2$ is the efficient coolant. 
However, there is an important difference;  
in the H atomic cooling case, 
the accretion rate onto the protostar is $\sim 0.1~\msunyr$, 
which is much higher than that in the ordinary Pop III case ($\sim 10^{-3}~\msunyr$).
This is due to the high temperature in the atomic cooling cloud 
($\simeq 8000$~K) because the accretion rate 
is set by the temperature in the star-forming cloud as 
$\dot M_{\rm acc} \sim 10^{-3}~\msunyr
(T/600~{\rm K})^{1.5}$ (Shu 1977; Stahler et al. 1986). 

This rapid accretion with $\sim 0.1~\msunyr$ 
drastically changes the protostellar evolution.
Figure \ref{fig:history} shows the evolution of the radii 
of accreting protostars at different accretion rates.
In the ordinary Pop III protostar case ($\mdot \simeq 10^{-3}~\msunyr$), 
after the so-called {\it adiabatic-accretion phase}, 
where adiabatic heat input expands the star gradually 
with mass and the protostar starts to contract 
by losing its entropy via radiative diffusion 
(the {\it Kelvin-Helmholtz contraction}) until
the nuclear ignition occurs at the center.
The protostar reaches the zero-age main sequence (ZAMS)
stage at this point
(Stahler, Palla \& Salpeter 1986; Omukai \& Palla 2001, 2003).
On the other hand, with the accretion rate as high as $\mdot \ga 0.1~\msunyr$, 
the protostar continues 
expanding without the KH contraction as recently 
found by Hosokawa, Omukai \& Yorke (2012, hereafter HOY12)
(see Figure \ref{fig:history}).
In such a star, while most of the interior material 
contracts, the outermost layer significantly swells up 
like a red-giant star (''{\it supergiant protostar}'' phase).
This is because the outer layer absorbs a part of the outward 
heat flux and obtains a very high specific entropy.
Also in this case the contraction at the center ceases
with the hydrogen ignition, but the envelope continuously 
expands with the increase of stellar mass.

%----------------------------------------------------------------------------%

If rapid accretion at $\mdot \ga 0.1~\msunyr$ is maintained,
the stellar mass exceeds $10^5~\msun$ within its lifetime. 
Such SMSs are general-relativistically unstable 
(e.g., Zel'dovich \& Novikov 1971; Shapiro \& Teukolsky 1983) 
and collapse as a whole to a BH (Shibata \& Shapiro 2002), 
which can be a seed for the SMBHs residing in the early universe ($z\ga 7$).
With the stellar mass increasing, however, the stars become 
more radiation-pressure dominated and approach a marginally 
stable state.
This may induce pulsational instability of the massive stars and 
result in mass-loss from the surface.
If this mass loss surpasses the accretion onto the star, 
the stellar growth will be terminated at that point.
To see whether the SMS formations are indeed possible in spite of 
such mass loss, we examine the pulsational stability 
of the supergiant protostars in this paper.

%----------------------------------------------------------------------------%

Pulsational stability of non-accreting Pop III stars has been 
studied by 
Baraffe, Heger \& Woosley (2001) and Sonoi \& Umeda (2012) in 
the range $120~\msun \leq M_{\ast} \leq 3 \times 10^3~\msun$.
They showed that those stars are unstable against pulsation caused 
by the nuclear burning 
(the so-called $\epsilon$ mechanism), and that the resulting
mass-loss rate is $\dot M_{\rm loss}\ga 10^{-5}~\msunyr$.
Gamgami (2007) studied this mass-loss process from the Pop III stars 
using 
spherically symmetric hydrodynamical simulations, and showed that
pulsation accelerates the surface material to the escape velocity 
and causes eruptive mass-loss for $M_{\ast}\ga 500~\msun$.
On the other hand, the Pop I red-giant stars are known to be
pulsationally unstable by the opacity-driven mechanism 
(the so-called $\kappa$ mechanism, e.g., Li \& Gong 1994; 
Heger et al. 1997), and the typical mass-loss rate is 
$\sim 10^{-5}~\msunyr$ (Yoon \& Cantiello 2010).
With the Pop III composition while
having similar structure to the Pop I 
red-giants, the supergiant protostars can also be pulsationally unstable.

In this paper, we study their stability by 
performing the linear stability analysis
for the mass range $M_{\ast} \la 10^3~\msun$, which has been 
calculated by HOY12. By estimating the mass-loss rate, 
we discuss whether supergiant protostars grow 
via accretion despite the pulsation-driven mass loss.

%---------------------------------------------------------------------------%

The organization of this paper is as follows.
In Section 2, we introduce the method for the linear stability analysis
against the pulsation and for the estimation of the mass-loss rates for unstable stars.
In Section 3, we present our results and explain how the stability
changes with the different stellar masses and accretion rates.
Finally, in Section 4, we summarize our study and present our discussions. 
In Appendix, we describe the details 
(the basic equations and boundary conditions)
of the linear stability analysis.

%%%%%%%%%%%%%%%%%%%%%%%%%%%%%
\section{Stability analysis}
%%%%%%%%%%%%%%%%%%%%%%%%%%%%%

\begin{figure}
\begin{center}
\includegraphics[height=60mm,width=80mm]{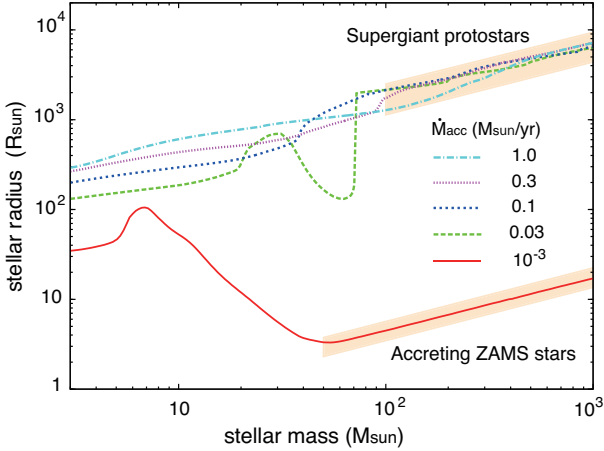}
\end{center}
\caption{Evolution of the protostellar radius with various accretion rates
$\dot M_{\rm acc} = 10^{-3}$, $0.03$, $0.1$,  $0.3$, and 
$1.0~\msunyr$ (taken from HOY12 with some modifications).
In this paper, we analyze the pulsational stability of the stars located 
in the shaded zones;  
accreting ZAMS stars for $\mdot = 10^{-3}~\msunyr$, and
supergiant protostars for $\mdot \ga 10^{-3}~\msunyr$.}
\label{fig:history}
\end{figure}

We study the pulsational stability of protostars growing
at constant accretion rates $\mdot = 10^{-3}$, 0.03, 0.1, 0.3, 
and $1.0~\msunyr$, whose structures have been numerically calculated 
in our previous work (HOY12).
Figure \ref{fig:history} presents the evolution of the stellar 
radii for these rates. 
We apply the linear stability analysis
(see Appendix for the details)
to stellar models either in the ZAMS (for $\mdot = 10^{-3} \msunyr$) 
or supergiant protostar (for higher accretion rates), indicated by 
the shaded areas in Figure~\ref{fig:history}.
We consider the perturbations proportional to $e^{i\sigma t}$,
where $\sigma = \sigma _{\rm R}+i\sigma _{\rm I}$ is the 
eigen frequency, $\sigma _{\rm R}$ the frequency of the pulsation,
and $|\sigma _{\rm I}|$ the growing or damping rate of the pulsation depending on 
its sign; the stars are stable (respectively unstable) if $\sigma_{\rm I} > 0$ 
(respectively $\sigma_{\rm I} < 0$).
According to previous studies, massive main-sequence Pop III stars
are unstable only under the radial perturbations 
(Baraffe et al. 2001; Sonoi \& Umeda 2012). 
We thus consider only the radial mode, hereafter, at which 
the supergiant protostars are also expected to be the most unstable.

%-------------------------------------------------------------------------%
%%% Fig.1 %%%
%--------------------------------------------------------------------------%

A useful quantity in the stability diagnosis is
the work integral $W$ (e.g., Cox 1980; Unno et al. 1989),
\begin{equation}
W(M_r)=\frac{\pi}{\sigma _{\rm R}}\int ^{M_r}_0
\Re \left[ 
\frac{\delta T^*}{T}
\left( \delta \epsilon - \frac{d}{dM_r}\delta L_{\rm rad} \right)
\right] dM_r,
\label{eq:work}
\end{equation}
where $M_r$ is the enclosed mass, $T$ is the temperature, 
$\epsilon$ is the nuclear energy generation rate per unit mass, 
$L_{\rm rad}$ is the radiative luminosity, and
the symbols with $\delta$ represent the Lagrange perturbations, 
where symbol $\Re$ denotes the real part of the quantity indicated in the bracket.
The work integral has the physical meaning of the pulsation 
energy gained inside $M_r$ in a single period.
If the sign of the work integral is positive 
at the stellar surface, i.e., $W(M_{\ast})>0$,
the stars gain kinetic energy in each period and are unstable. 
The pulsation amplitude increases during the growth timescale 
of the instability $\sigma_{\rm I}^{-1}$.
If $W(M_{\ast})<0$, on the other hand, the pulsation damps 
inside the stars and are stable.
The first term in the bracket on the right-hand side of equation (\ref{eq:work}), proportional to $\delta \epsilon$, 
represents the driving of instability 
by the nuclear burning (i.e., the $\epsilon$ mechanism).
The second term is related to the radiative energy transport.
In most cases, the radiative diffusion damps the pulsation, 
and thus the second term contributes to the stabilization.
However, in the surface layer where the opacity changes remarkably,
the energy flux transported via radiation can be absorbed and 
be converted into the pulsation energy by the $\kappa$ mechanism.
The growth (or damping) rate of the pulsation per single period 
$\eta \equiv - \sigma_{\rm I}/\sigma_{\rm R}$ is written as
\begin{equation}
\eta \equiv -\frac{\sigma _{\rm I}}{\sigma_{\rm R}}
= \frac{W(M_{\ast})}{4\pi E_{\rm W}},
\label {eq:relation}
\end{equation}
(see Cox 1980; Unno et al. 1989), 
where
\begin{equation}
E_{\rm W}=\frac{\sigma _{\rm R}^2}{2}\int ^{M_{\ast}}_0|\xi _r|^2 dM_r
\label{eq:energy}
\end{equation}
is the pulsation energy,
and $\xi_r$ is the radial displacement of fluid elements from 
their equilibrium positions.

%--------------------------------------------------------------------------%

In stars which are unstable under linear perturbations, 
the pulsation amplitude will grow to the non-linear regime.
Such a strong pulsation is expected to cause a mass loss 
from the stellar surface (Appenzeller 1970a, b; Papaloizou 1973a, b).
Appenzeller (1970a, b) studied the non-linear growth of 
the radial-pulsation instability for Pop I massive stars 
of $M_\ast = 130$ and $270~\msun$ using 
one-dimensional hydrodynamical calculations.
He showed that after the pulsation enters the non-linear regime, 
the surface velocity reaches the speed of sound and 
weak shocks emerge just inside the photosphere.
The shocks recurrently propagate outward and accelerate 
the gas in the surface layer (e.g., Lamers \& Cassinelli 1999), 
generating mass shells exceeeding the escape velocity which is then 
lost from the star.

%---------------------------------------------------------------------------%

In this paper, we evaluate the mass-loss rate following
Baraffe et al. (2001) and Sonoi \& Umeda (2012).
As shown by Appenzeller (1970a, b), outflows are launched when the 
pulsation velocity at the surface reaches the speed of sound $c_{\rm s}$.
At this moment, the pulsation amplitude at the surface is
\begin{equation}
\xi _{r, \rm surf}=\frac{c_{\rm s}}{\sigma _{\rm R}}.
\label{eq:normalization}
\end{equation}
Using this, we can estimate the pulsation energy $E_{\rm W}$ as well as the work integral $W$.
Assuming that all the pulsation energy is converted into the kinetic energy of the outflows,
the mass-loss rate $\dot M_{\rm loss}$ can be obtained 
from the energy conservation:
\begin{equation}
\frac{\dot M_{\rm loss}}{2} v_{\rm esc}^2 
= \frac{\sigma _{\rm R}}{2\pi} W(M_{\ast}) 
= -2 \sigma_{\rm I} E_{\rm W},
\label{eq:energy_conserve}
\end{equation}
where $v_{\rm esc}=(2GM_{\ast}/R_{\ast})^{1/2}$ is the escape velocity.

%---------------------------------------------------------------------------%

Note that the assumption of energy conservation above is not 
always valid 
because some pulsation energy can be lost by radiative dissipation. 
In fact, Papaloizou (1973a, b) obtains a mass-loss rate lower than that of 
Appenzeller(1970a, b) by one order of magnitude by including this effect.
The mass-loss rate derived below can thus be regarded as a conservative upper limit.

%%%%%%%%%%%%%%%%%%
\section{Results}
%%%%%%%%%%%%%%%%%%

In this Section, we describe the results for two different regimes 
of the accretion rate separately: (a) high accretion-rate cases ($\mdot \geq 0.03~\msunyr$),
where the accreting stars become supergiant protostars, 
and (b) a low accretion-rate case ($\mdot = 10^{-3}~\msunyr$), 
where it reaches the ordinary ZAMS.
The high-rate regime corresponds to the cases of the SMS formations,
while the lower rates are expected in the ordinary Pop III 
star formation. 
The latter results are presented for comparison 
with the previous studies (Baraffe et al. 2001; Sonoi \& Umeda 2012).
Since we have found that accreting protostars are unstable only for 
the radial mode without nodes (fundamental mode or ``F-mode''), 
we present the results for the F-mode below.

\subsection{High accretion-rate cases 
($\mdot \geq 0.03~\msunyr$): supergiant protostars}

We see here the high accretion-rate cases $\mdot \geq 0.03~\msunyr$, 
where the protostars grow in mass through the
supergiant-protostar phase (Figure~\ref{fig:history}).
We first explain the case with the highest accretion rate
$\mdot = 1.0~\msunyr$ and then the cases with the lower rates.

\subsubsection{Highest Accretion-Rate Case 
($\dot M_{\rm acc} = 1.0~\msunyr$)}
\label{sssec:highest}
\begin{figure}
\begin{center}
\includegraphics[height=60mm,width=75mm]{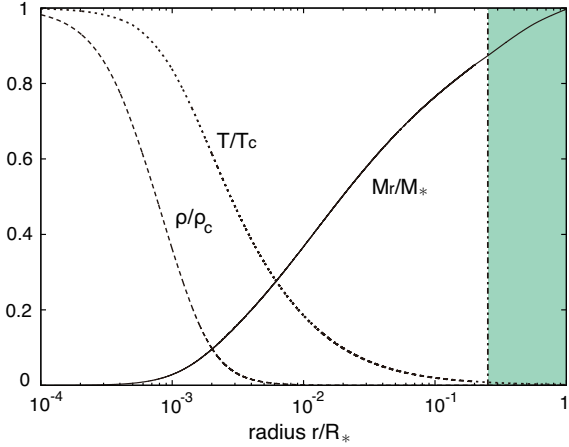}
\end{center}
\caption{The interior structure of the accreting
$10^3~\msun$ protostar with $\dot M_{\rm acc}=1.0~\msunyr$ 
as a function of the relative radius $r/R_{\ast}$.
The lines present the radial profiles of 
the enclosed mass (solid), density (dashed), 
and temperature (dotted), respectively.
The enclosed mass is normalized by the stellar mass and others are 
normalized by their central values; 
$\rho _{\rm c}=0.2$ g cm$^{-3}$ and $T _{\rm c}=2.1\times 10^7$ K.
The vertical line at $r/R_{\ast} \simeq 0.25$ denotes the 
inner boundary of the convective envelope.}
\label{fig:interior_SMPS}
\end{figure}

Figure~\ref{fig:interior_SMPS} shows the stellar interior structure 
when the stellar mass reaches $10^3~\msun$ with $\mdot = 1.0~\msunyr$.
No convective core develops in the interior since the
hydrogen burning has not yet started.
The star instead consists of a radiative core and an outer 
convective layer.
Although the convective layer only constitutes 10\% of the stellar mass,
and the remaining 90\% is the radiative core, it covers a large portion of 
the radius.
This structure consisting of the central core and bloated envelope 
is similar to that of red-giant stars. 
With $\dot M_{\rm acc} = 1.0~\msunyr$, the protostar reaches 
this structure at $M_\ast \ga 200~\msun$.
In this evolutionary stage, the stellar luminosity is close to
the Eddington value ($L_{\ast} \simeq L_{\rm Edd} \propto M_{\ast}$), 
and the effective temperature remains almost constant at 
$T_{\rm eff} \simeq 5000$~K due to the strong temperature-dependence 
of the H$^-$ bound-free opacity.
With these two conditions, the mass-radius relationship 
of the supergiant protostars is analytically written as
\begin{equation}
R_{\ast}\simeq 8.2\times 10^3~{\rm R}_{\sun}
\left(\frac{M_{\ast}}{10^3~\msun}\right)^{1/2},
\label{eq:MR_SMPS}
\end{equation}
which well agrees with the numerical results (HOY12).

%---------------------------------------------------------------------------%
%%% Fig.2 %%%
%---------------------------------------------------------------------------%

\begin{figure}
\begin{center}
\includegraphics[height=60mm,width=80mm]{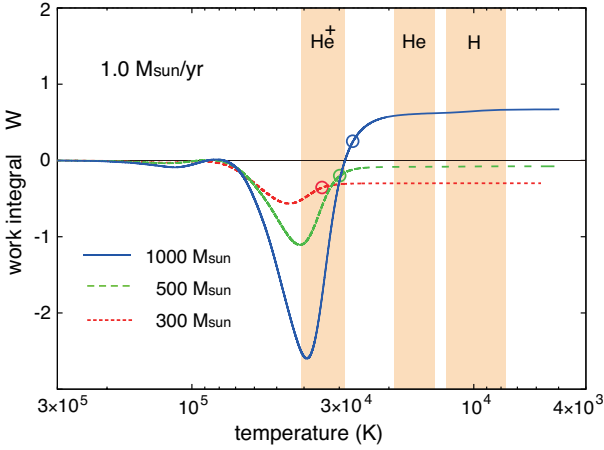}
\end{center}
\caption{
Radial distributions of the work integral $W$ 
(in an arbitrary unit) near the stellar
surface ($T < 3 \times 10^5$~K) for 
$\dot M_{\rm acc}=1.0~\msunyr$.
The lines represent the $M_{\ast}=300$ (dotted), 
$500$ (dashed), and $10^3~\msun$ (solid) stars.
The shaded zones denote the ionization layers of 
He$^+$, He, and H from left to right.
Open circles mark the transition points, where the thermal timescale
is equal to the dynamical timescale, $t_{\rm th} = t_{\rm dyn}$.   
}
\label{fig:work_temp_SMPS}
\end{figure}

Figure~\ref{fig:work_temp_SMPS} shows the spatial distributions of 
the work integral for the radial F-mode at the stellar masses of 
$300$, $500$, and $10^3~\msun$.
The work integral $W$ changes remarkably near 
the surface ($\la  3\times 10^5~{\rm K}$), 
in particular, around $4 \times 10^4$~K, due to 
a opacity bump by the He$^+$ ionization.
At $300$ and $500~\msun$, the work integrals are negative 
at the surface and the stars are stable.
At $M_{\ast} = 10^3~\msun$, on the other hand, the work integral 
at the stellar surface is positive, i.e.,
the protostar is pulsationally unstable by the $\kappa$ mechanism 
excited in the He$^+$ ionization layer.

%----------------------------------------------------------------------------%

All the work integrals shown in Figure~\ref{fig:work_temp_SMPS}
are constant in the outer H$^0$ and He$^0$ ionization layers
since the radiative energy transport is efficient enough there. 
The $\kappa$ mechanism neither excite nor damp the pulsation.
This can be seen by comparing the following two timescales,  
the cooling time in the layer outside a radius $r$ in the 
unperturbed state (thermal timescale; e.g., Sonoi \& Shibahashi 2011)
\begin{equation}
t_{\rm th} \equiv \frac{\int ^{R_{\ast}}_r 4\pi c_P T\rho r^2dr}{L},
\label{eq:tth}
\end{equation}
and the period of the pulsation
\begin{equation}
t_{\rm dyn} \equiv  \frac{2 \pi}{\sigma_{\rm R}}.
\label{eq:tdyn}
\end{equation}
The open circles in the Figure~\ref{fig:work_temp_SMPS} 
indicate the transition points where the two timescales equal each other ($t_{\rm th} = t_{\rm dyn}$).
Outside this point, $t_{\rm th}$ is shorter than $t_{\rm dyn}$ 
as the density and the specific heat decrease outward.
We call this region where $t_{\rm th} < t_{\rm dyn}$ 
as the {\it non-adiabatic zone}.
The variation of the work integral is almost zero there 
because the entropy is rapidly dissipated during a pulsation period.
Thus, the surface value of the work integral, which determines 
the pulsational stability of the star, is fixed at the transition point to the
non-adiabatic zone, where $t_{\rm th} = t_{\rm dyn}$.

%---------------------------------------------------------------------------%

As seen in Figure~\ref{fig:work_temp_SMPS}, 
the surface value of the work integral $W(M_{\ast})$ increases 
with the stellar mass and eventually becomes positive for $\ga 500~\msun$:
the protostar becomes pulsationally unstable.
Since the work integral grows in the He$^+$ ionization layer inside the transition 
point but remains constant outside.
This increase of the surface value $W(M_{\ast})$ with the stellar mass 
can be understood by the concomitant outward-shift of the transition point, 
which in turn can be explained by comparing the two timescales;
\begin{equation}
t_{\rm th} \propto \frac{R_{\ast}^3}{L_{\ast}},
\label{eq:tth_power}
\end{equation}
and 
\begin{equation}
t_{\rm dyn} \propto \sqrt{\frac{R_{\ast}^3}{M_{\ast}}}
\label{eq:tdyn_power}
\end{equation}
near the surface.
In deriving the equation (\ref{eq:tth_power}),
we used the fact that the term $c_P T \rho$ in equation
(\ref{eq:tth}) changes only slightly for $T < 4\times 10^5$~K in the range
$10^2~\msun \la M_{\ast} \la 10^3~\msun$.
Note that the dynamical timescale (equation~\ref{eq:tdyn_power}) has 
the same dependence as the free-fall timescale of the star. 
Eliminating $R_{\ast}$ and $L_{\ast}$ in equations (\ref{eq:tth_power})
and (\ref{eq:tdyn_power}) with equation (\ref{eq:MR_SMPS}) and
$L_{\ast} \simeq L_{\rm Edd}$, we obtain:
\begin{equation}
\frac{t_{\rm dyn}}{t_{\rm th}} \propto M_{\ast}^{-1/4}, 
\label{eq:therm_dyn}
\end{equation}
the thermal timescale becomes longer with respect to the dynamical timescale 
near the surface with increasing stellar mass.
In other words, the surface layer becomes more adiabatic:
the non-adiabatic zone on the surface layer becomes thinner and 
the transition point moves outward 
as seen in Figure~\ref{fig:work_temp_SMPS}.
As a result, the surface value of the work integral increases and 
the stars become more unstable as the stellar mass increase.

%---------------------------------------------------------------------%
%%% Fig.3 %%%

%-----------------------------------------------------------------------%

\begin{figure}
\begin{center}
\includegraphics[height=60mm,width=80mm]{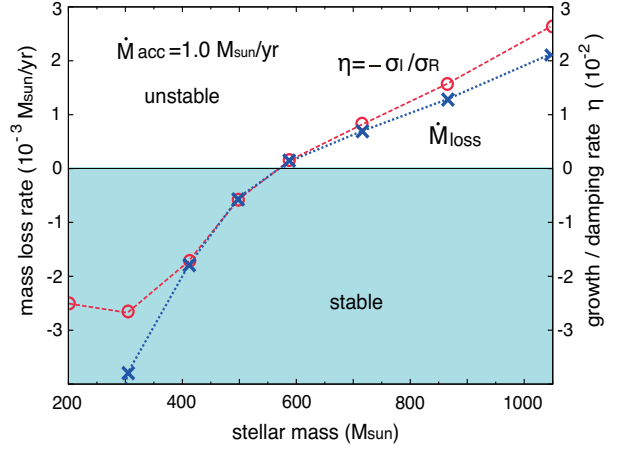}
\end{center}
\caption{The growth/damping rate 
$\eta(=-\sigma _{\rm I}/\sigma_{\rm  R})$ and the mass-loss rate 
$\dot M_{\rm loss}$
as a function of stellar mass for $\dot M_{\rm acc} =1.0~\msunyr$.
The left (right) vertical axis shows $\dot M_{\rm loss}$ in unit of 
$10^{-3}~\msunyr$ ($\eta$ in unit of $10^{-2}$, respectively).
In the shaded area the star is stable against 
the radial pulsation, i.e., $\eta<0$.
}
\label{fig:mass_loss_1e0}
\end{figure}

The growth rate of the pulsation $\eta$ and the resulting 
mass-loss rate 
$\dot M_{\rm loss}$ are shown in Figure~\ref{fig:mass_loss_1e0}  
as a function of the stellar mass.
At $M_{\ast} \simeq 600~\msun$, the star becomes 
pulsationally unstable 
and the mass loss rate increases with the stellar mass
thereafter.
At $M_{\ast} \simeq 10^3~\msun$, 
the mass-loss rate reaches $2 \times 10^{-3}~\msunyr$, 
two orders of magnitude higher than that in the ZAMS case
with accretion rate $\mdot = 10^{-3}~\msunyr$ 
(see Sec. 3.2 below).
This is, however, still lower than the accretion rate 
by a factor of $500$. 
In the case with spherical symmetry, therefore,
pulsation-driven outflow would be completely quenched by 
the rapid accretion. With some angular momentum,
the accretion onto the star proceeds mostly through a circumstellar disk.
In this case, the outflow escapes unhindered in the polar directions 
where the stellar surface is not covered by the accreting flow.
We thus expect that the supergiant protostar loses 
some material via bipolar pulsation-driven outflows, while 
simultaneously growing in mass due to a more rapid accretion 
from the disk.

%-------------------------------------------------------------------------%
%%% Fig.4 %%%
%--------------------------------------------------------------------------%
%--------------------------------------------------------------------------%
%%% Fig.5 %%%
\begin{figure}
\begin{center}
\includegraphics[height=60mm,width=80mm]{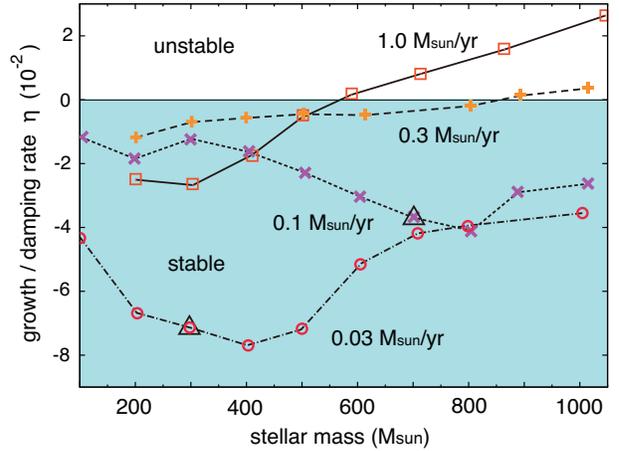}
\end{center}
\caption{The growth/damping rate 
$\eta =-\sigma_{\rm I} /\sigma _{\rm R}$ as a function of 
stellar mass for $\dot M_{\rm acc}=1.0$ (solid), $0.3$ (long-dashed), 
$0.1$ (short-dashed), $0.03~\msunyr$ (dash-dotted), respectively. 
In the shaded area the star is stable against 
the radial pulsation ($\eta<0$).
The symbols on the lines show the models for which
we analyze the stability. 
Large open triangles on the cases with $0.1$ and 
$0.03~\msunyr$ indicate the onset of the hydrogen burning.
}
\label{fig:loss_SMPS}
\end{figure}
%-------------------------------------------------------------------------%

\subsubsection{Variation with different accretion rates}

Next we see the cases with lower accretion rate $0.03-0.3~\msunyr$.
Figure~\ref{fig:loss_SMPS} presents the growth
rate $\eta$ for the radial F-mode in these cases as functions of the stellar mass. 
Roughly speaking, 
at a given stellar mass, the growth rate $\eta$ is higher for higher accretion rates.
In our analysis, stars are unstable (i.e. $\eta >0$) only in the two highest accretion rate cases,:   
those with $1.0~\msunyr$ for $M_{\ast} \ga 600~\msun$ and with $0.3~\msunyr$ for 
$M_{\ast} \ga 900~\msun$.

%----------------------------------------------------------------------------%

\begin{figure}
\begin{center}
\includegraphics[height=60mm,width=80mm]{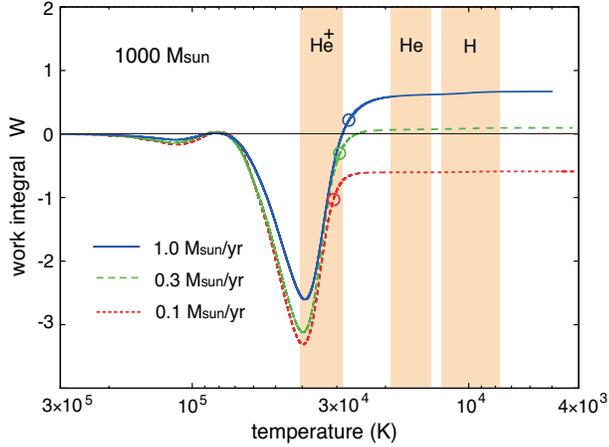}
\end{center}
\caption{The same as Figure~\ref{fig:work_temp_SMPS}, but for the 
stellar models at $10^3~\msun$ with three 
different accretion rates 
$\dot M_{\rm acc}=1.0~\msunyr$ (solid), 
$0.3~\msunyr$ (dashed), and $0.1~\msunyr$ (dotted). 
}
\label{fig:work_temp_SMPS_1}
\end{figure}

\begin{figure}
\begin{center}
\includegraphics[height=110mm,width=80mm]{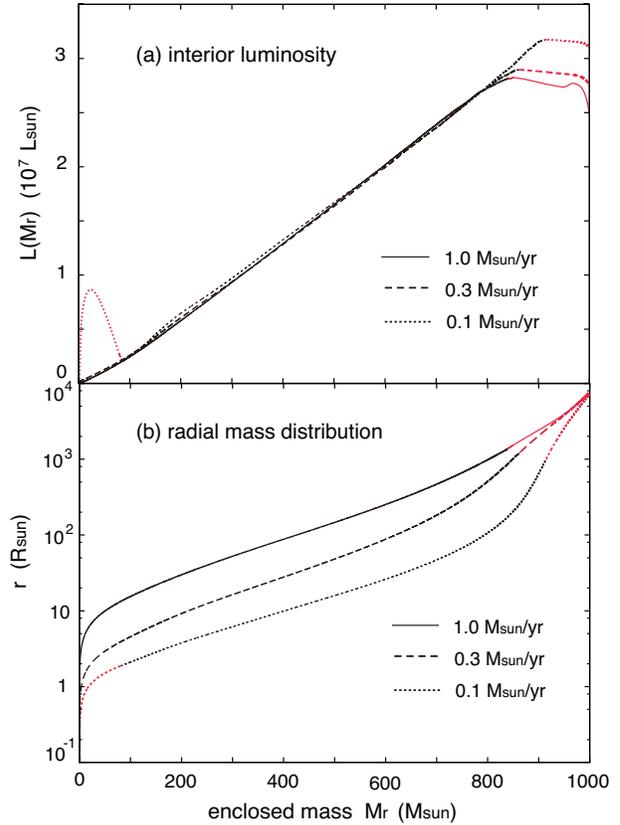}
\end{center}
\caption{Comparison of the interior structure of $10^3~\msun$
protostars with different accretion rates.
The panels (a) and (b) present the radial distributions of
the luminosity and enclosed mass, respectively.
In the both panels, the red portions denote the convective zones.
}
\label{fig:mass_rho_lum}
\end{figure}

This tendency of instability toward higher accretion rates
can be understood again from the outward-shift of the 
transition point between the adiabatic and non-adiabatic zones 
inside the He$^+$ ionization layer, 
which makes the surface value of the work integral 
$W(M_{\ast})$ higher (see Section \ref{sssec:highest}). 
The behavior of work integral $W$ is shown 
in Figure~\ref{fig:work_temp_SMPS_1} 
for three accretion rates of 0.1, 0.3, and $1.0~\msunyr$ 
at $M_{\ast}=10^3~\msun$.
%----------------------------------------------------------------------------%
%%% Fig.6 %%%
The ratio of the timescales $t_{\rm dyn}/t_{\rm th}$ depends
on the accretion rate $\dot M_{\rm acc}$ only through 
the stellar surface luminosity $L_*$ 
(see equations \ref{eq:MR_SMPS}, \ref{eq:tth}, and \ref{eq:tdyn}
and note that the stellar radius is independent of $\dot M_{\rm acc}$). 
As shown in Figure~\ref{fig:mass_rho_lum} (a) 
the surface luminosity $L_{\ast}$ and the ratio $t_{\rm dyn}/t_{\rm th}$ is lower 
for higher $\dot M_{\rm acc}$. 
In other words, the surface region becomes more adiabatic and 
the transition point moves closer to the surface at 
higher $\dot M_{\rm acc}$, which makes the surface value of 
work integral higher as well.

%-----------------------------------------------------------------%

The above relation of $L_{\ast}$ and $\dot M_{\rm acc}$ can 
be understood in the following way.
With lower $\dot M_{\rm acc}$, the central part of the star 
has longer time to lose its entropy and the star takes a 
more centrally concentrated structure at a given stellar mass 
maintaining the same stellar radius
(see Figure~\ref{fig:mass_rho_lum} b).
As the radiative energy transport is efficient in the 
inner hot and dense part, such a star has larger 
radiative core.
Since the luminosity grows proportionally to the enclosed mass $M_r$ 
inside the radiative core but remains roughly constant 
outside (see Figure~\ref{fig:mass_rho_lum} a), 
the large radiative core at low $\dot M_{\rm acc}$ results in 
high value of surface luminosity $L_{\ast}$. 

%--------------------------------------------------------------------------%
%%% Fig.7 %%%

%---------------------------------------------------------------------------%

Although the overall behavior of the growth rate 
$\eta$ shown in Figure~\ref{fig:loss_SMPS} can be understood
with the above considerations, $\eta$ evolves in a somewhat
complicated way at $\dot M_{\rm acc}=0.03$ and $0.1~\msunyr$; 
i.e., $\eta$ decreases with mass early in the 
evolution. The reason is as follows.
As seen above, 
a supergiant protostar becomes more centrally concentrated 
and thus more stable (i.e., lower $\eta$) with the increasing mass.
At the same time, however, there is also a destabilization effect with mass 
which is due to the shrinking of the non-adiabatic layer of the surface, 
as discussed in Sec.~\ref{sssec:highest}.
These two effects compete each other.
With the highest accretion rates of $\mdot = 0.3$ and 
$1.0~\msunyr$, the destabilization is more important, 
while in the cases with low accretion rate of $\mdot = 0.03$ 
and $0.1~\msunyr$, the stabilization first dominates
until the onset of hydrogen burning, after which 
the central concentration remains 
almost constant and the stabilizing effect no longer operates.
Thus, at this point $\eta$ begins to increase as 
seen in Figure~\ref{fig:loss_SMPS}.

\subsection{Lowest accretion-rate case
($\dot M_{\rm acc}=10^{-3}~\msunyr$ ): Accreting ZAMS stars}
\label{ssec:lowacc}

%---------------------------------------------------------------------------%
%%% Fig.8 %%%
\begin{figure}
\begin{center}
\includegraphics[height=60mm,width=72mm]{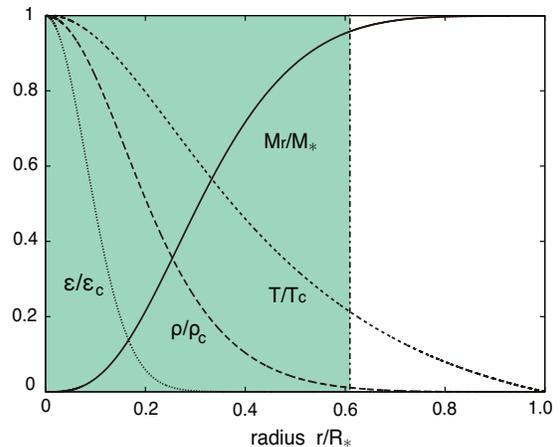}
\end{center}
\caption{The same as Figure 2, but for the 
$10^3~\msun$ protostar with $\dot M_{\rm acc}=10^{-3}~\msunyr$. 
The energy production rate due to the nuclear burning is 
also shown (dotted).
The enclosed mass is normalized by the stellar mass, and others are 
by their central values:
$\rho_{\rm c} = 11$ g cm$^{-3}$, $T _{\rm c}=1.3\times 10^8$ K,
and $\epsilon_{\rm c} = 6.9 \times 10^5$ erg s$^{-1}$ g$^{-1}$.
The vertical line at $r/R_{\ast} \simeq 0.6$ denotes the outer boundary 
of the convective core.}
\label{fig:interior_MS}
\end{figure}
%--------------------------------------------------------------------------%
%--------------------------------------------------------------------------%
%%% Fig.9 %%%
\begin{figure}
\begin{center}
\includegraphics[height=60mm,width=72mm]{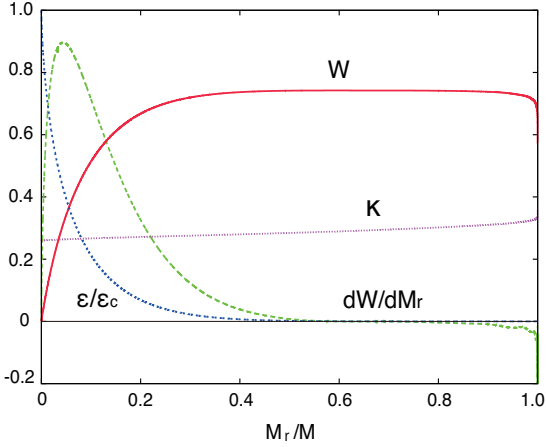}
\end{center}
\caption{
Radial distributions of several quantities within the
$M_{\ast}=10^3~\msun$ star with $\mdot = 10^{-3}~\msunyr$.
The work integral $W$ (solid line) and its derivative $dW/dM_r$
(long-dashed line) are presented in arbitrary units.
The nuclear energy production rate $\epsilon$ (normalized by the central value) and the opacity $\kappa$ (in cm$^2$ g$^{-1}$)
are plotted with the short-dashed and dotted lines, respectively. 
}
\label{fig:work_temp_MS}
\end{figure}
%---------------------------------------------------------------------------%

Next, we will see the lowest accretion-rate case 
in our calculation with
$\dot M_{\rm acc} = 10^{-3}~\msunyr$, where
the protostar reaches the ZAMS at $M_\ast \simeq 50~\msun$ 
after the KH contraction (e.g., Omukai \& Palla 2003).

%--------------------------------------------------------------------------%

Figure~\ref{fig:interior_MS} presents the interior structure 
of the protostar $M_{\ast} = 10^3~\msun$: 
the radial profiles of the mass, the temperature,
the density, and the nuclear energy production rate.
The central hydrogen burning via CN-cycle renders the 
95\% of the stellar mass (covering 60\% in radius) to be convective.
The vertical line (dot-dashed) in Figure~\ref{fig:interior_MS}
indicates the boundary between the convective core and 
the outer radiative envelope.

%-------------------------------------------------------------------------%
 
Figure~\ref{fig:work_temp_MS} shows the radial distributions
of the work integral $W$, its derivative $dW/dM_r$, 
its nuclear-energy production rate $\epsilon$, 
and its opacity $\kappa$ within this star.
The $\epsilon$-mechanism drives the pulsational instability and thus 
the work integral $W$ increases inside the convective core.
On the other hand, the pulsation is slightly damped 
(i.e. $dW/dM_r<0$) in the radiative layer because of the energy dissipation. 
The $\kappa$ mechanism does not work as the opacity 
is almost constant in the envelope due to high surface 
temperature ($\sim 10^5$K).
The small mass inside the surface region cannot totally 
damp the pulsation excited by the $\epsilon$ mechanism. 
The star is thus unstable, i.e., $W(M_{\ast}) > 0$.

%---------------------------------------------------------------------------%

\begin{figure}
\begin{center}
\includegraphics[height=60mm,width=80mm]{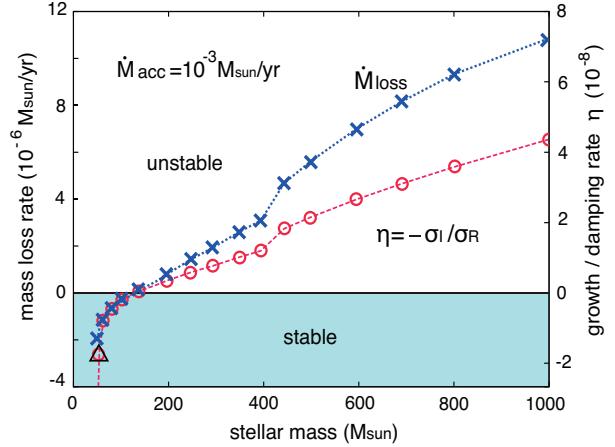}
\end{center}
\caption{The growth/damping rate 
$\eta(=-\sigma _{\rm I}/\sigma_{\rm  R})$, and the mass loss rate 
$\dot M_{\rm loss}$ as a function of stellar mass for 
$\dot M_{\rm acc} =10^{-3}~\msunyr$.
The left (right) vertical axis shows $\dot M_{\rm loss}$ 
in unit of $10^{-6}~\msunyr$ ($\eta$ in unit of $10^{-8}$, respectively).
The protostar is stable ($\eta<0$) against 
the radial pulsation (F-mode) in the shaded area.
Open triangles indicate the onset of the hydrogen burning.}
\label{fig:loss_1e3}
\end{figure}

Figure~\ref{fig:loss_1e3} presents 
the growth rate of the pulsation $\eta$ and the 
resulting mass-loss rate $\dot M_{\rm loss}$ as a function of 
the stellar mass.
After hydrogen ignition at $M_{\ast}\simeq 50~\msun$, 
the star remains stable until $140~\msun$ when the
stabilization by radiative damping overcomes 
the pulsation by the $\epsilon$ mechanism.
As the stellar mass increases, however, the star becomes 
more radiation-pressure dominated and 
the average adiabatic exponent of the star 
$\Gamma_1\equiv (\partial \ln p/\partial \ln \rho )_S$ approaches 
the marginal gravitational stability value of $4/3$.
Consequently, the pulsation becomes increasingly strong 
in the central convective core and exceeds the 
radiative damping effect
(e.g., Cox 1980; Shapiro \& Teukolsky 1983).
The star becomes unstable at $140~\msun$ and 
the growth rate of the pulsation increases thereafter.
The mass-loss rate is typically
$\dot M_{\rm loss}\simeq 10^{-6}-10^{-5}~\msunyr$ 
(the dotted line in Figure~\ref{fig:loss_1e3}).
Since this mass-loss rate is lower than 
the accretion rate $\dot M_{\rm acc}= 10^{-3}~\msunyr$,  
the stellar growth via accretion would not be 
prevented by the pulsation-driven mass-loss as in the 
supergiant protostar cases.

%----------------------------------------------------------------------%

Baraffe et al. (2001) and Sonoi \& Umeda (2012) also studied
the pulsational instability of non-accreting massive Pop III stars.
For example, Sonoi \& Umeda (2012) estimate the growth rate
of the pulsation and mass-loss rate for a 
$500~\msun$ star as $\eta =3.02\times 10^{-8}$ and 
$\dot M_{\rm loss} = 2.0 \times 10^{-5}~\msunyr$, respectively.
Although their growth rate agrees well with our results, 
their mass-loss rate is higher than ours by a factor of four.
This difference in mass-loss rates comes from the 
different values of the surface density between 
the accreting and non-accreting stars.
With accretion, the surface density 
is higher than that without accretion.
In this case, the gas pressure is relatively higher 
than the radiation pressure, i.e., higher 
$\beta \equiv p_{\rm gas}/p_{\rm tot}$. 
This results in a lower sound velocity 
$c_{\rm s} = \sqrt{\Gamma_1p/\rho}~(\propto \beta ^{-1/2})$
at the surface and 
a lower pulsation amplitude at the onset of the mass
loss, which is proportional to the speed of sound 
(equation~\ref{eq:normalization}). 
Therefore, the pulsation energy $E_{\rm W}$ and the 
mass-loss rate become lower in the accreting case 
than in the non-accreting case.

%---------------------------------------------------------------------------
%%% Fig.10 %%%
%----------------------------------------------------------------------------

%%%%%%%%%%%%%%%%%%%%%%%%%%%%%%%%%%%%%
\section{Conclusion and Discussion}
\label{sec:concl}
%%%%%%%%%%%%%%%%%%%%%%%%%%%%%%%%%%%%%

In this paper, we have studied the pulsational stability
of primordial protostars growing via very rapid accretion, 
($\mdot \sim 0.1~\msunyr$),
through the method of the linear perturbation analysis, which is 
expected in the case of supermassive star formation 
in the early universe.
We have evaluated mass-loss rate if the protostar is 
pulsationally unstable and examined whether the mass loss 
is strong enough to prevent the stellar growth via the accretion.
We focused on early stellar evolution of 
$M_\ast \la 10^3~\msun$, which has been studied in our recent 
work (HOY12). Our results are summarized as follows.

%-------------------------------------------------------------------------------%

First, we have studied the high accretion-rate cases with 
$\mdot \ga 0.03~\msunyr$, where the protostar has the a contracting core 
and a bloated envelope similar to a giant star 
({\it supergiant protostar}; HOY12).
With low effective temperature $T_{\rm eff} \simeq 5000$~K, 
the supergiant protostar has the H and He ionization 
layers within its envelope.
We have found that although pulsation is driven due to blocking of  
radiative flux at the opacity bump from the 
He$^+$ ionization (the so-called $\kappa$ mechanism), 
the supergiant protostars are pulsationally 
unstable only with the highest accretion rate  
$\simeq 1.0~\msunyr$ we studied.
In the lower accretion-rate cases, the protostars are stable at least 
until $M_{\ast}\simeq 10^3~\msun$.
Even in the most unstable cases, the mass-loss rates
are typically $\sim 10^{-3}~\msunyr$, which are lower than
their accretion rates by more than two orders of magnitude. 
We thus conclude that the mass loss driven by pulsation
does not prevent the growth of the supergiant protostar via rapid accretion.

%-------------------------------------------------------------------------------------%

Next, for comparison with the previous studies, 
we have analyzed a lower accretion-rate case with 
$\mdot = 10^{-3}~\msunyr$, which is expected in the 
ordinary Pop III star formation.
In this case, the protostar reaches the ZAMS at 
$M_{\ast} \simeq 50~\msun$ after the KH contraction stage
(e.g., Omukai \& Palla 2001, 2003).
We have found that the protostars are unstable by the
$\epsilon$ mechanism in the range $M_\ast \ga 140~\msun$,
where a large part of the stellar interior is 
radiation-pressure dominated.
Estimated mass-loss rate $10^{-6}-10^{-5}~\msunyr$
is roughly consistent with the previous results for non-accreting stars
(Baraffe et al. 2001; Sonoi \& Umeda 2012) although smaller by few factors 
because of the difference in the surface density due to the accretion.

%------------------------------------------------------------------------------------%

In this paper, we have limited our analysis to 
the mass range $M_{\ast}\la10^3~\msun$ due to the lack 
of stellar data in the higher mass range. 
We here speculate the later evolution based on the current results. 
Further studies on the protostellar evolution 
for $M_{\ast} > 10^3~\msun$ as well as 
on its pulsational stability are thus awaited. 
If we linearly extrapolate the mass-loss rate 
in the case of $\mdot = 1.0~\msunyr$ 
shown in Figure~\ref{fig:mass_loss_1e0}
to higher mass range, 
\begin{equation}
\dot M_{\rm loss}\sim 5.0\times 10^{-4}
\left(\frac{M_{\ast}}{100~\msun} -6\right)~\msunyr;
\label{eq:mass_loss_rate}
\end{equation}
the mass loss catches up with the accretion at 
$M_{\ast} \simeq 2\times 10^5~\msun$.
At this point, 
the growth of the protostar via accretion possibly halts 
and the final mass is set. 
However, because of our assumption that all the
pulsation energy is converted to the kinetic energy of the outflows, 
the mass-loss rate by equation  (\ref{eq:mass_loss_rate})
should be regarded as an upper limit 
(e.g., Papaloizou 1973a, b).
Furthermore, the mass loss of the Pop I red-giant stars are usually  
driven by the radiation pressure exerted on dust grains
formed in the cool envelope (e.g., Willson 2000).
Without the dust as in our case, acceleration of the outflows 
could be more inefficient.
We expect that, with such rapid accretion,  
the final stellar mass can exceed $\sim 10^5~\msun$ 
despite the pulsation-driven mass loss.

Although mass-loss rate is much lower than the accretion rate 
until $M_\ast = 10^3~\msun$ as studied in this paper, 
these values could be comparable to $M_\ast \ga 10^5~\msun$. 
Since the accretion of gas with some angular momentum 
onto the star proceeds via a
circumstellar disk, the outflows would escape most easily in the 
polar regions, where the density is relatively low.
The dynamical interaction between the inflows and outflows needs to 
be studied in detail to determine the exact value 
of the stellar final mass.

Our estimate of the mass-loss rate is based on the previous works 
(e.g., Appenzeller 1970a, b; Papaloizou 1973a, b), in which 
the non-linear development of pulsation 
for non-accreting main-sequence stars is studied 
numerically.
For the accreting stars, however, 
we have a very limited knowledge on the non-linear 
behavior of pulsation (e.g., Gamgami 2007).
More detailed studies on this issue by radiative hydrodynamical 
simulations is awaited.

%---------------------------------------------------------------------------------%

So far, we have only considered stars forming 
from the metal-free ($Z=0$) gas.
However, SMSs could potentially be formed from the gas slightly
polluted with heavy elements, 
if that is below the critical amount $Z_{\rm cr}$, 
i.e., $\sim 10^{-3}~Z_{\sun}$ without dust grains,
or $\sim 10^{-5}~Z_{\sun}$ with dust grains 
(Omukai et al. 2008; Inayoshi \& Omukai 2012).
Metal enrichment lowers the central temperature of a star 
by enhancing the energy production efficiency by nuclear fusion 
and also creates another opacity bump near the surface which 
make the pulsation stronger via 
the $\epsilon$ and $\kappa$ mechanisms, respectively.
However, for the $\kappa$ mechanism, for which the supergiant 
protostars are unstable, this effects becomes important only with 
metallicity higher than $2\times 10^{-3}~Z_{\sun}$ in 
the case of non-accreting stars (Baraffe et al. 2001), 
which is higher than the critical value $Z_{\rm cr}$. 
We thus speculate that  
even if small amount of metals below $Z_{\rm cr}$ are present, 
the pulsational stability of supergiant protostars 
should be similar to the zero-metallicity case studied above. 

%---------------------------------------------------------------------------------%

Finally, we discuss the validity of the frozen-in approximation 
of convective energy flux used in our analysis (see Appendix A), 
where perturbations of the convective flux is neglected.
At the present time, this is a widely-used approximation due to our
limited knowledge on the interaction between 
the convective and pulsational motions.
Although some other models including this effect have been
proposed (e.g., Unno 1967; Gough 1977; Unno et al. 1989; Dupret et al. 2005),
they rely on the still-developing time-dependent convection theories, 
which require different assumptions depending on modeling, 
beyond the classical mixing-length theory (B{\"o}hm-Vitense 1958).
Recent results by Penev, Barranco \& Sasselov (2009) and 
Shiode, Quataert \& Arras (2012), who studied this interaction 
numerically, showed that the convective damping weakens the pulsation by 
the $\epsilon$ mechanism, but does not influence through 
the $\kappa$ mechanism.
Therefore, protostars with modest accretion rate 
$\dot M_{\rm acc}\simeq 10^{-3}~\msunyr$, which are unstable by 
the $\epsilon$ mechanism (Sec.~\ref{ssec:lowacc}), can be somewhat stabilized 
by this convective damping.
On the other hand, we speculate that this would not significantly
affect the pulsation in supergiant protostars, which is 
driven by the $\kappa$ mechanism.
\\

\section*{Acknowledgments}

We would like to thank Takashi Nakamura for his continuous 
encouragement, Takafumi Sonoi and Kei Tanaka for fruitful discussions, 
and Shunsuke Katayama for improving the manuscript.
This work is in part supported by the Grants-in-Aid by the Ministry 
of Education, Culture, and Science of Japan (23$\cdot $838 KI; 2168407 
and 21244021 KO).

\appendix

%%%%%%%%%%%%%%%%%%%%%%%%%
\section{Linear Perturbation Analysis Method}
%%%%%%%%%%%%%%%%%%%%%%%%%

In this appendix, we describe our method of the linear 
perturbation analysis of the stellar pulsation stability
(e.g., Cox 1980; Unno et al. 1989).
The basic equations governing the stellar structure are 
\begin{equation}
\frac{\partial \rho}{\partial t}+\nabla \cdot (\rho {\bf v})=0,
\label{eq:continuous}
\end{equation}
\begin{equation}
\frac{\partial {\bf v}}{\partial t}+({\bf v}\cdot \nabla ){\bf v}
=-\frac{1}{\rho}\nabla p-\nabla \Phi,
\label{eq:euler}
\end{equation}
\begin{equation}
\nabla ^2 \Phi=4\pi G\rho,
\label{eq:poisson}
\end{equation}
\begin{equation}
T\left[\frac{\partial S}{\partial t}+({\bf v}\cdot \nabla )S\right]
=\epsilon -\frac{1}{\rho}\nabla \cdot {\bf F},
\label{eq:entropy}
\end{equation}
\begin{equation}
{\bf F}_{\rm rad}=-\frac{4ac}{3\kappa \rho}T^3\nabla T,
\label{eq:radiation}
\end{equation}
where $\rho$ is the density, ${\bf v}$ the velocity, $p$ the pressure, 
$\Phi$ the gravitational potential, $T$ the temperature, 
$S$ the specific entropy, $\epsilon$ the nuclear-energy generation rate 
per unit mass, $\kappa$ the opacity, 
${\bf F}$ the total energy flux, which is the sum of the radiative
flux ${\bf F}_{\rm rad}$ and convective flux ${\bf F}_{\rm conv}$,
$G$ the gravitational constant, $c$ the speed of light, 
and $a$ the radiation constant. 
The radial mode, i.e., perturbations
which radially oscillate with an eigen frequency $\sigma$, 
is studied.

We consider the radial displacement of fluid elements from the 
equilibrium positions in the form $\xi_r(r,t) \equiv \xi_r(r)e^{i\sigma t}$.
We define the resulting Euler perturbation of a physical quantity $Q$ 
as $Q'\equiv Q(r,t)-Q_0(r,t)$, where $Q_0$ is the value in the
unperturbed state.
We also use the Lagrange perturbation
$\delta Q \equiv Q(r+\xi_r,t)-Q_0(r,t)$ for some physical quantities.
The linearized equations (\ref{eq:continuous})-(\ref{eq:radiation})
with the perturbations $Q'(r,t)=Q'(r)e^{i\sigma t}$ and 
$\delta Q(r,t)=\delta Q(r)e^{i\sigma t}$ are written as
\begin{equation}
\frac{1}{r^2}\frac{d}{dr}(r^2 \xi_r)-\frac{g}{c_{\rm s}^2}\xi _r +
\frac{p'}{\rho c_{\rm s}^2}=v_{\rm T}\frac{\delta S}{c_{\rm P}},
\label{eq:pertub2}
\end{equation}
\begin{equation}
\frac{1}{\rho}\frac{dp '}{dr}+\frac{g}{\rho c_{\rm s}^2}p'+(N^2-\sigma ^2)\xi _r
+\frac{d\Phi '}{dr}=g v_{\rm T}\frac{\delta S}{c_{\rm P}},
\label{eq:pertub1}
\end{equation}
\begin{equation}
\frac{1}{r^2}\frac{d}{dr}\left( r^2\frac{d\Phi '}{dr}\right)
-4\pi G\rho \left( \frac{p'}{\rho c_{\rm s}^2}+\frac{N^2}{g}\xi_r \right)
=-4\pi G\rho v_{\rm T}\frac{\delta S}{c_{\rm P}},
\label{eq:pertub3}
\end{equation}
\begin{equation}
i\sigma T\delta S=\delta \epsilon -\frac{d\delta L_{\rm rad}}{dM_r},
\label{eq:pertub4}
\end{equation}
\begin{equation}
\frac{\delta L_{\rm rad}}{L_{\rm rad}}=-\frac{\delta \kappa }{\kappa}+4\frac{\delta T}{T}
+4\frac{\xi _r}{r}
+\frac{d(\frac{\delta T}{T})/d\ln r}{d\ln T/d\ln r},
\label{eq:pertub5}
\end{equation}
where $c_{\rm s}~(=\sqrt{\Gamma_1p/\rho})$ is the sound velocity, 
$\Gamma_1=(\partial \ln p/\partial \ln \rho )_S$ the adiabatic exponent,
$r$ the radius, $g$ the gravitational acceleration, 
$L_{\rm rad}$ the radiative luminosity,
$M_r$ the enclosed mass, $c_{\rm P}=T(\partial S/\partial T)_p$ 
the isobaric specific heat,
$v_{\rm T}\equiv -(\partial \ln \rho/\partial \ln T)_p$, 
and $N^2\equiv -g(d\ln \rho/dr+g/c_{\rm s}^2)$ 
the Brunt-V$\ddot{\rm a}$is$\ddot{\rm a}$r$\ddot{\rm a}$ frequency.
In the above equations, for simplicity,  a physical quantity ``$Q$'' indicates
its value in the unperturbed state instead of $Q_0$.

Note that, in equations (\ref{eq:pertub4}) and (\ref{eq:pertub5}), 
we ignore the perturbation of the convective energy flux, i.e., $\delta {\bf F}_{\rm conv}=0$.
This so-called ``frozen-in'' approximation 
has been widely used in analyzing the pulsational 
stability of stars (Baraffe et al. 2001 and Sonoi \& Umeda 2012). 
To facilitate the comparison between their results,
we also adopt this assumption here
(see Section~\ref{sec:concl} for more discussions).

From equations (\ref{eq:pertub2}) and (\ref{eq:pertub3}) and the regularity of $\Phi '$ at the center,
\begin{equation}
\frac{d\Phi '}{dr}+4\pi G\rho \xi_r=0.
\end{equation}
Eliminating the term $d\Phi '/dr$ in equations
(\ref{eq:pertub1})-(\ref{eq:pertub5}) with this relation, we obtain
four linear ordinary first-order differential equations
for four variables $\xi_r$, $p'$, $\delta S$, and $\delta L_{\rm rad}$.
Here, we impose the following boundary conditions:
\begin{equation}
\frac{d}{dr}\left(\frac{\xi_r}{r}\right)=0,
~~\frac{d}{dr}\left(\frac{\delta L_{\rm rad}}{L_{\rm rad}}\right)=0~~~~~(r=0),
\end{equation}
\begin{equation}
\frac{d}{dr}\left(\frac{\delta p}{p}\right)=0~~~~~(r=R_{\ast}),
\end{equation}
from the regularity of the perturbations at the center
and surface, and
\begin{equation}
\frac{\delta F_{{\rm rad},r}}{F_{{\rm rad},r}}=4\frac{\delta T}{T}~~~~~~(r=R_{\ast}),
\end{equation}
which guarantees outward propagation of the energy flux 
at the surface
(e.g., Cox 1980; Saio, Winget \& Robinson 1983).
In this system of the differential equations and boundary 
conditions, the normalization of the variables
$\xi_r$, $p'$, $\delta S$, $\delta L_{\rm rad}$ still remains 
as a degree of freedom.
We solve the system as an eigenvalue problem by
formally introducing a differential equation for 
the eigenvalue $\sigma$,
\begin{equation}
\frac{d\sigma}{dr}=0.
\end{equation}
The whole system here is the five first-order differential equations for
$\xi_r$, $p'$, $\delta S$, $\delta L_{\rm rad}$, and $\sigma$
with the four boundary conditions and one normalization condition.
We set the arbitrary normalization condition at the surface, 
$\xi _r(r=R_{\ast})=R_{\ast}$. 
We obtain numerical solutions of the eigen functions and 
eigenvalue using the relaxation method (e.g., Unno et al. 1989).

\bsp

\label{lastpage}

\end{document}